\documentclass[aip,cha,amsmath, amssymb,reprint]{revtex4-1}

\usepackage{graphicx}
\usepackage{dcolumn}
\usepackage{bm}
\usepackage{xcolor}

\DeclareMathAlphabet{\mathsfsl}{OT1}{cmss}{m}{sl}
\newcommand{\vect}[1]{\boldsymbol{#1}}

\newcommand{\mi}{\mathrm{i}}
\newcommand{\rd}{\mathrm{d}}

\newcommand{\oh}{\frac{1}{2}}
\newcommand{\oq}{\frac{1}{4}}

\newcommand{\mydef}
        {\stackrel{\mathrm{def}}{=}}

\newcommand{\EASD}{\langle || \Delta \vect{r}(\tau) ||^2 \rangle_\mathrm{E}}
\newcommand{\TASD}{\langle || \Delta\vect{r}(\tau) ||^2 \rangle_\mathrm{T}}
\newcommand{\EATASD}{\langle \langle || \Delta\vect{r}(\tau) ||^2 \rangle_\mathrm{T} \rangle_\mathrm{E}}

\begin{document}

\title{Normal and anomalous random walks of 2-d solitons}
\author{Jaime Cisternas}
\email{jecisternas@miuandes.cl}
\affiliation{Complex Systems Group, Facultad de Ingenier\'{\i}a y Ciencias Aplicadas,
Universidad de los Andes, Monse\~nor Alvaro del Portillo 12455, Las Condes, Santiago, Chile}
\author{Tony Albers}
\affiliation{Institute of Physics, Chemnitz University of Technology, 09107 Chemnitz, Germany}
\author{G\"unter Radons}
\affiliation{Institute of Physics, Chemnitz University of Technology, 09107 Chemnitz, Germany}

\date{\today}

\begin{abstract}
Solitons, which describe the propagation of concentrated beams of light through
nonlinear media, can exhibit a variety of behaviors as a result of the intrinsic dissipation, diffraction, and the nonlinear effects.
One of these phenomena, modeled by the complex Ginzburg-Landau equation, are chaotic explosions, transient enlargements of the soliton
that may induce random transversal displacements, which in the long run lead to a random walk
of the soliton center. As we show in this work, the transition from non-moving to moving solitons
is not a simple bifurcation but includes a sequence of normal and anomalous random walks.
We analyze their statistics with the distribution of generalized diffusivities,
a novel approach that has been used successfully for characterizing anomalous diffusion.
\end{abstract}

\maketitle


\begin{quotation}
Powerful coherent laser light in crystals or fibers can suffer nonlinear interactions
that prevent its divergence and foster stable propagation as a single entity: a soliton.
In this theoretical work, we study solitons that have been found in experimental setups
where energy is continuously injected to overcome dissipation. Using a simple mathematical model
for the dynamics of the electromagnetic waves in the nonlinear media, we analyze the long
term behavior of the center of the soliton (projected into a transversal plane)
and characterize the statistics of the trajectories
using mathematical concepts developed for random walks.
\end{quotation}

\section{Introduction}

Localized waves that travel through space without major changes in their shape
appear in several physical phenomena\cite{DescalziClerc,Liehr}, for instance, hydrodynamics\cite{KBS88,NAC90},
optics\cite{TSW97,USM03}, and surface reactions\cite{RJVE91,BBK15}.
Their existence is explained as a balance between nonlinearity, dispersion, diffraction, and energy fluxes\cite{AkhmedievAnkiewicz}.
The name `dissipative soliton' is reserved for those structures that
share some properties with `classical solitons' of integrable models (for instance the nonlinear Schr\"odinger equation),
although they are not conservative.
These dissipative solitons are either stationary, oscillating, or chaotic and may move with more or less fixed shape and velocity.

One of the peculiar phenomena of dissipative solitons are `explosions'\cite{SAA00,AST01,AS04}:
transient enlargements of amplitude and size of the soliton induced by the chaotic internal dynamics
without the need of external perturbations. These short-lived explosions do not destroy the
soliton.
Experimental evidence of explosions of solitons has been obtained in several experiments
involving the propagation of strong light pulses
in Ti:sapphire mode-locked lasers \cite{CSA02} and
in double-clad ytterbium-doped fibers \cite{GA12,RBE15,RBE16,LLY16,LLX16}.

Recent work of the authors\cite{CCDB12,CDAR16, CAR17} focussed on trajectories of solitons
that can be described as random walks.
The position of the soliton, as a result of spatial jumps with random orientations induced by explosions,
has a probability density that spreads diffusively in the plane.
We have shown\cite{CDAR16} how this diffusive behavior can be anomalous, i.e. the mean-squared displacement of the location of the soliton
grows nonlinearly with time.

Anomalous diffusion has been observed in a large variety of natural and artificial systems,
and in the last decades, a sophisticated mathematical framework has been developed \cite{MW65,KlafterSokolov,MJC14,KlagesRadonsSokolov}.
The study of anomalous diffusion has made a
fascinating progress, both theoretically and experimentally, and benefited from
numerous reports outside the domain of Physics.


This article is a continuation of Ref.~\onlinecite{CDAR16} and has the following three objectives:
(i) Describe the transition between no diffusion (soliton exploding in-place) and normal diffusion as a single parameter is varied.
(ii) Understand the role of anomalous diffusion in the transition.
(iii) Use new measures to describe different regimes of diffusion, particularly the \emph{distribution of generalized diffusivities} introduced in Ref.~\onlinecite{AR13}
in extension of the concepts of Ref.~\onlinecite{bauer2011}.

\section{The model}

The basic model for the study of explosions of dissipative solitons is the complex cubic-quintic Ginzburg-Landau equation (CGL)\cite{ADA08,SAD08}:
\begin{equation}
\partial_t A = D \nabla_\perp^2 A + \mu A + \beta |A|^2 A + \gamma |A|^4 A
\label{cqgl2d}
\end{equation}
on a two-dimensional spatial domain and
subject to specified initial condition $A(x,y,0)$.
The operator $\nabla_\perp^2=\partial_x^2+\partial_y^2$ is the isotropic Laplacian in two dimensions.
Complex coefficients $D,\beta, \gamma$ are assumed constant and spatially uniform.
The real coefficient $\mu$ (distance to the onset of linear instability) is also constant and uniform and
is going to be used as our main bifurcation parameter.
$D$ represents diffraction and spectral filtering;
$\mu$ accounts for losses; $\beta$ represents cubic nonlinear gain and Kerr effect; and
$\gamma$ is the quintic nonlinear gain.

This model is relevant for wide aperture lasers, vertical external cavity semiconductor devices,
and multimode optical fibers made of erbium doped glass\cite{WRCW15,WCW17}.

There are wide regions of combinations of parameters $(\mu,D,\beta,\gamma)$ where solutions can be localized in the plane
with exponential tails that decay independently of the domain size $L$.

In this article, we used: $\beta=1+0.8\ \mi, \gamma=-0.1-0.6\ \mi, D=0.125+0.5\ \mi$.
We are interested in solutions that have shapes similar to a two-dimensional soliton.
As it was mentioned in the Introduction, these solitons exhibit different behaviors:
stationary, pulsating, or explosive.
Explosive solitons experience infinite sequences of symmetric and asymmetric explosions.
As it was found in Ref.~\onlinecite{CDAR16}, only the latter induce diffusive behavior
but the former are relevant for the statistical properties.




The two-dimensional CGL, Eq.~(\ref{cqgl2d}), was integrated from a localized initial condition using a split-step Fourier method.
With the exception of $\mu$, all other parameters were kept fixed in all the simulations reported here.

The simulations were carried out using a $256 \times 256$ spatial grid of size $dx \times dy = 0.2 \times 0.2$.
The time discretization was $dt=0.005$, and the runs typically involved $2 \times 10^7$ iterations,
so the total time was $T=10^5$, so several thousands of explosions were registered in each run.

We checked our results using a finer spatial grid $512 \times 512$
($dx$ was halved so $dt$ was divided by 4 to respect the numerical accuracy) that the results were not artefacts of the spatial or temporal discretizations.
The spatial power spectrum converged exponentially fast, so higher modes can be considered irrelevant.

Our codes were implemented on graphical processing units using the PyOpenCL library developed by Andreas Kl\"ockner \cite{pycuda}.

\section{Effective diffusion of solitons}

Spatial diffusion is induced by random asymmetric explosions of the soliton that take place intermittently as a result of its chaotic dynamics.
The sensitive dependence on initial conditions, typical for chaotic systems, implies that the
precise timing of the explosions, their symmetry, and their spatial directions seem to be random.

We have found\cite{CCDB12,CDAR16} that there are several regimes with significative differences.
First there is a regime for values of $\mu \le -0.40$, where all explosions are perfectly symmetric
and the location of the soliton remains stationary.
Then there is another regime for values of $\mu \ge -0.30$, where most explosions are asymmetric
and lead to a random walk of the location of the soliton that obeys the statistics of Brownian motion:
uncorrelated jumps of finite size and finite waiting times between the jumps.
But there is an intermediate range, $-0.40 < \mu < -0.30$, where one can find
for some range of values of $\mu$ a third alternative:
long sequences of symmetric explosions interspersed with some asymmetric explosions.
As the time between asymmetric explosions can be arbitrarily long in the sense that they follow a heavy-tailed distribution with infinite mean,
the resultant random walk has statistics different from Brownian motion and falls in the class of anomalous diffusion, more specifically, subdiffusion,
which can be modeled by a subdiffusive continuous time random walk (CTRW).

To quantify the differences between these regimes, we define the coordinates of the `center of mass' of the soliton:
\begin{equation*}
x_\mathrm{cm}(t) \mydef \frac{L}{2\pi} \mathrm{arg} \left( \int |A|^2 \exp\left( \frac{\mi 2\pi x}{L} \right) \, \rd x\, \rd y \right)
\end{equation*}
and analogously for $y_\mathrm{cm}(t)$.
These definitions respect the periodicity of the domain and can be used to define
the vector center of mass:
\begin{equation*}
\vect{r}(t) \mydef \left( \begin{array}{c} x_\mathrm{cm}(t) \\ y_\mathrm{cm}(t) \end{array}\right)
\end{equation*}
(in the following, we will drop the subscript `cm').

\begin{figure}
\begin{center}
\includegraphics[width=0.5\textwidth]{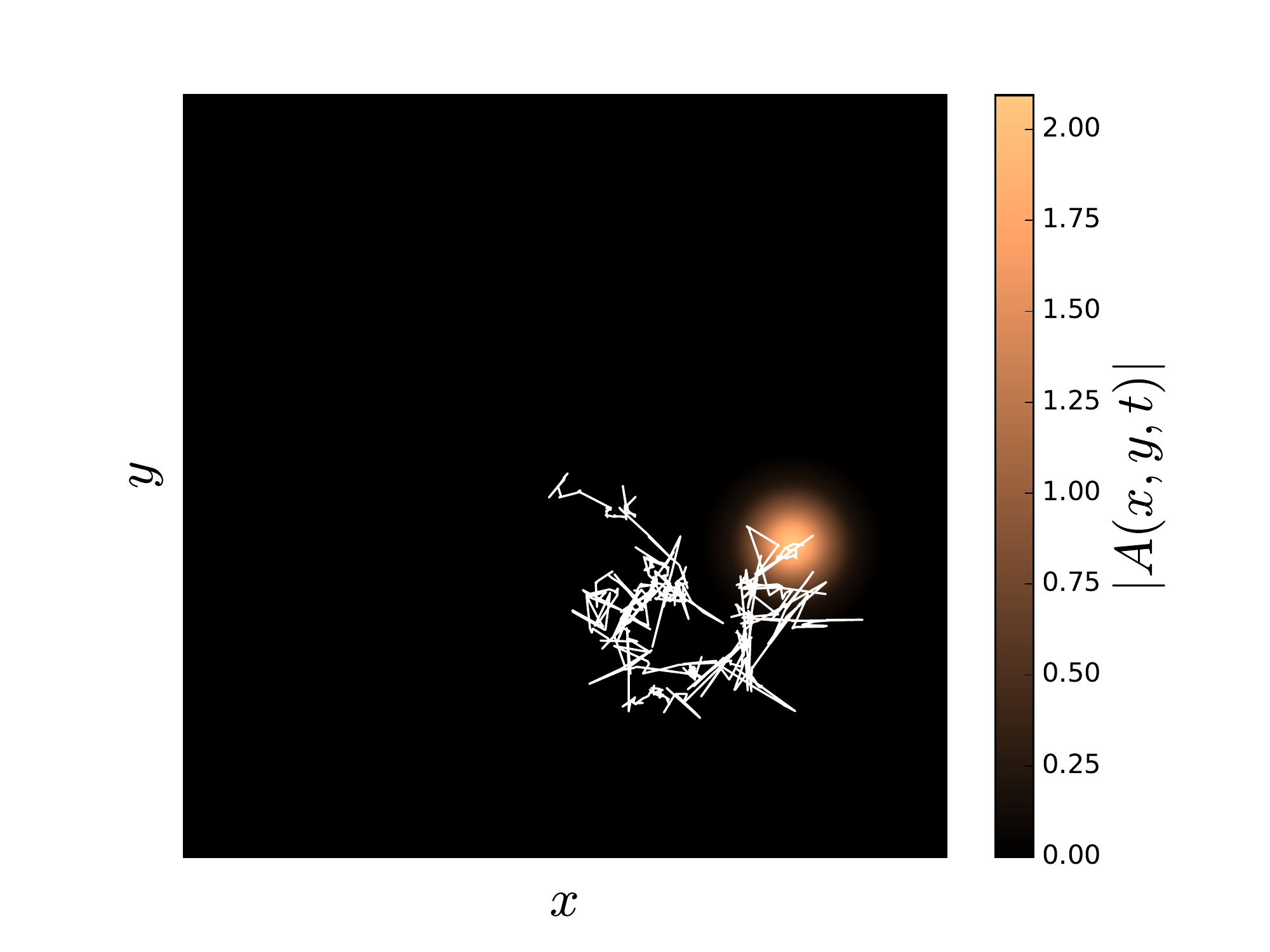}
\bigskip
\end{center}
\caption{The white trajectory is an example of the chaotic evolution of the center $(x(t), y(t))$ of the soliton over a time $T=10^5$
as a result of repeated asymmetric explosions that lead to spatial shifts in random directions.
The trajectory, initialized at the center of the square domain, is superimposed to a density plot of the absolute value of the soliton amplitude $A(x,y,t)$
at the final time $t=T$.
}
\label{fig:overview}
\end{figure}

In Fig.~\ref{fig:overview}, we plot the 2-d soliton and the trajectory of its center of mass.
It was generated from an initial condition, where we
added some small spatial noise to a Gaussian pulse,
by integrating
the deterministic law represented by Eq.~(\ref{cqgl2d}).
Asymmetric explosions induce random jumps.
As it was mentioned above, the
coexistence of symmetric and asymmetric explosions may lead to anomalous random walks
if the mean time between asymmetric explosions is infinite.

The basic quantity that is used to characterize diffusive motion is the mean-squared displacement (MSD),
which can be defined either as ensemble average or as time average.

\begin{figure}
\begin{center}
\includegraphics[width=0.5\textwidth]{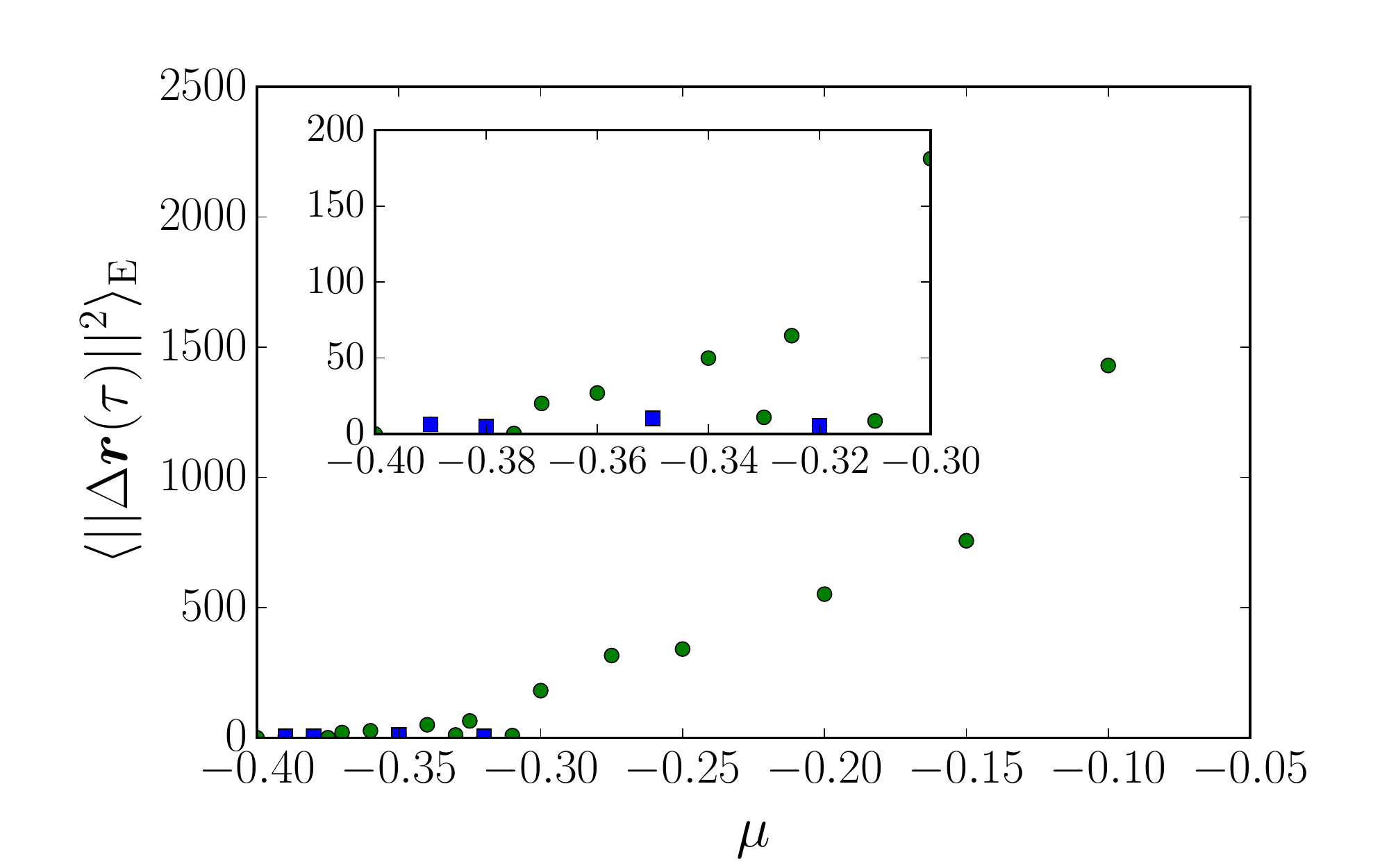}
\end{center}
\caption{Summary of behaviors of EASD defined by Eq.~(\ref{ensmsd})
evaluated at $\tau=10^4$.
For $\mu<-0.4$, the location of the soliton remains static.
For $\mu>-0.3$, the location of the soliton describes a normal random walk.
In the intermediate range, as indicated in the small inset, diffusive motion
alternates with regimes of no diffusion and with anomalous diffusion (indicated by blue squares).
}
\label{fig:easd}
\end{figure}

The ensemble-averaged squared displacement (EASD) is defined by
\begin{equation}
\EASD \mydef \lim_{N \rightarrow \infty} \frac{1}{N} \sum_{i=1}^N || \vect{r}_i(\tau) ||^2
\label{ensmsd}
\end{equation}
assuming $\vect{r}_i(0)=0$.
In Fig.~\ref{fig:easd}, we plot such averages obtained from ensembles of $N=64$ independent realizations for a variety of values of $\mu$
at time $\tau=10^4$.
The plot shows, as we sweep the interval with increments of size $\Delta \mu=0.01$, a general trend of larger EASD for larger $\mu$, but
in the intermediate regime for $\mu=-0.380,-0.350,-0.320$, there are some points with very small displacements.
They may correspond to static solitons or some other form of diffusion.

The EASD asympotically increases with a power law,
\begin{equation}
\EASD  \simeq \langle D_{\alpha}\rangle_{\mathrm{E}}\,\tau^\alpha\quad(\tau\rightarrow\infty), 
\end{equation}
where $\langle D_{\alpha}\rangle_{\mathrm{E}}$ is the generalized diffusion coefficient.
The exponent $\alpha$ seems to be 1 for most values of $\mu$, that indicates normal diffusion and is consistent with the shape of the trajectories.
But, as we will see, in some cases inside the intermediate regime, 
there are indications of anomalous diffusion, $\alpha \ne 1$.
Although the exponent $\alpha$ is the primary signal of anomalous diffusion,
finite simulation times and the limited number of trajectories make its estimation difficult.


The time-averaged squared displacement (TASD) of an individual trajectory is defined by
\begin{equation}
\TASD \mydef \frac{1}{T-\tau} \int_0^{T-\tau} || \vect{r}(t+\tau)-\vect{r}(t) ||^2 \, \rd t.
\label{tamsd}
\end{equation}
This quantity is not necessarily `reproducible' even for long simulation times $T$, i.e.
independent realizations (from different initial conditions) may give very different values of $\TASD$.

\begin{figure}
\begin{center}
\includegraphics[width=0.5\textwidth]{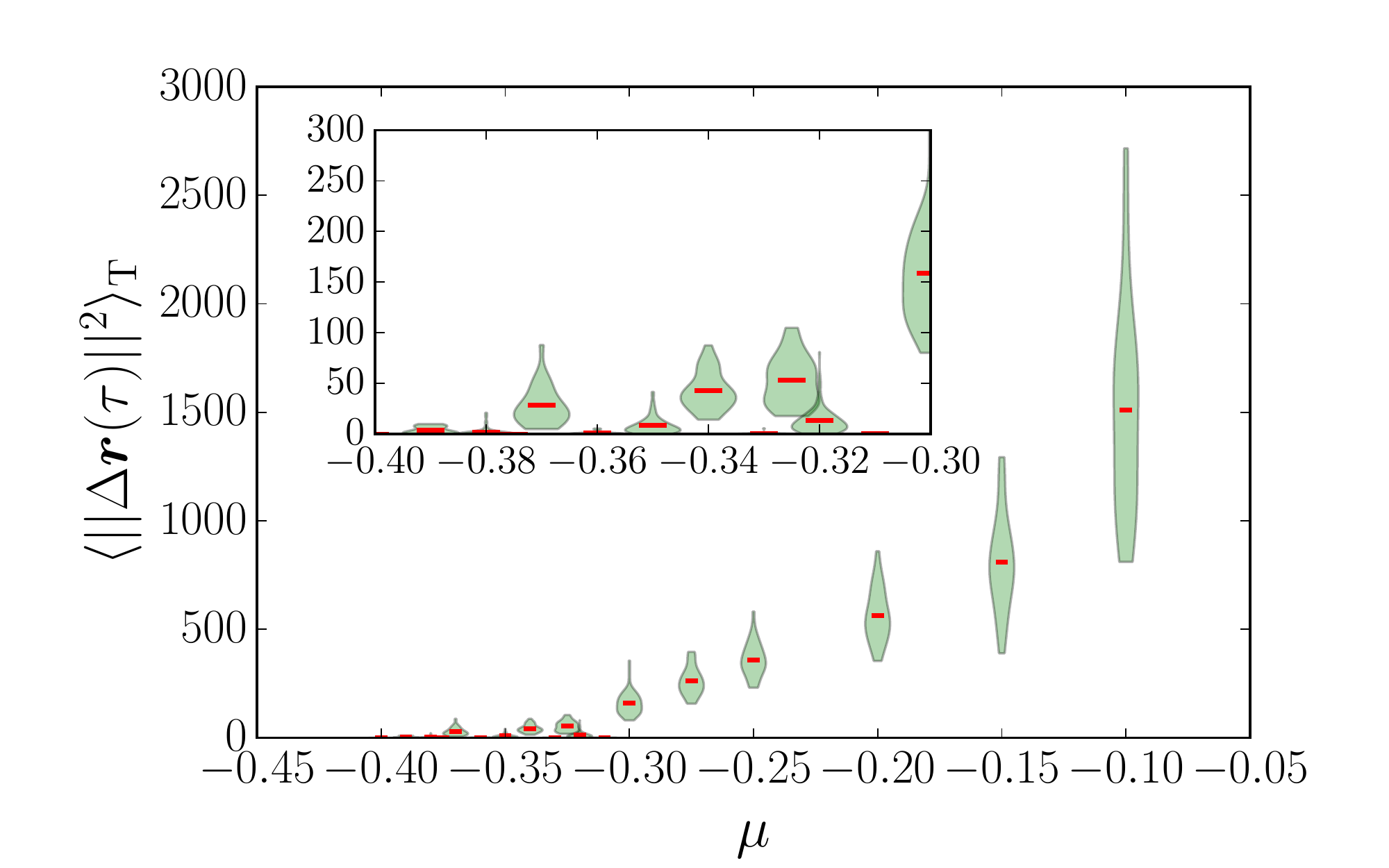}
\end{center}
\caption{Summary of behaviors of TASD defined by Eq.~(\ref{tamsd}) evaluated for $\tau=10^4, T=10^5$.
For each value of $\mu$, the vertical shape indicates the distribution of values that $\TASD$
takes over the realizations. Red horizontal bars indicate the ensemble average of $\TASD$.
Anomalous cases show distributions that are concentrated not far from zero.
}
\label{fig:tasd}
\end{figure}

The scaling of the TASD often turns out to be quite different from the EASD.
The TASD gives a result for each independent trajectory.
Our results for all $\mu$ indicate that for long $T$ and $\tau < T$, the TASD grows linearly with time,
\begin{equation}
\TASD \simeq \langle D_1 \rangle_\mathrm{T}\ \tau ~.
\end{equation}
Now as we will show, for finite $T$, the proportionality coefficient of the TASD is a random variable. 
In the case of normal difusion, there is a small scatter that gets smaller for larger values of $T$.
In the case of anomalous diffusion, this proportionality coefficient varies wildly from realization to realization (the values span several orders of magnitude) and these fluctuations remain, even for infinite $T$.
Fig.~\ref{fig:tasd} shows the large width of these distributions.

The last plot also shows (with horizontal bars) the ensemble averages of the TASD:
$$\EATASD  =  \lim_{N \rightarrow \infty} \frac{1}{N} \sum_{i=1}^N \langle||\Delta\vect{r}_i(\tau)||^2\rangle_{\mathrm{T}} .$$


Comparing Fig.~\ref{fig:easd} with Fig.~\ref{fig:tasd} and the observed scaling relations (using finite $T$ and $\tau<T$),
there are some indications that for some values of $\mu$ :
\begin{equation*}
\EATASD \ne \EASD ~.
\end{equation*}
The last inequality receives the name `weak ergodicity breaking' and leads to several counter-intuitive effects\cite{MJC14}.




\begin{table}
\begin{center}
\begin{tabular}{ccc}
$\mu$ & ~~~ & Behavior \\
\hline
-0.40 & & no diffusion \\
-0.39 & & anomalous \\
-0.38 & & anomalous \\
-0.37 & & normal \\
-0.36 & & no diffusion \\
-0.35 & & anomalous \\
-0.34 & & normal \\
-0.33 & & no diffusion \\
-0.32 & & not clear \\
-0.31 & & no diffusion \\
-0.30 & & normal \\
\hline
\end{tabular}
\end{center}
\caption{Different behaviours reveal transition between non-moving soliton to walking solitons as the parameter $\mu$ is increased.
In the intermediate range, small windows of anomalous diffusion are interspersed with non-moving and normal diffusion.
}
\label{tab:summary}
\end{table}


In the following, we focus our attention into two very different scenarios.
We will show that the exponent $\alpha$ extracted from EASD is hard to measure accurately,
and it is far better to look at the distribution of TASD,
in particular, the distribution of $\langle D_1 \rangle_\mathrm{T}$.

\begin{figure*}
\begin{center}
\includegraphics[width=\textwidth]{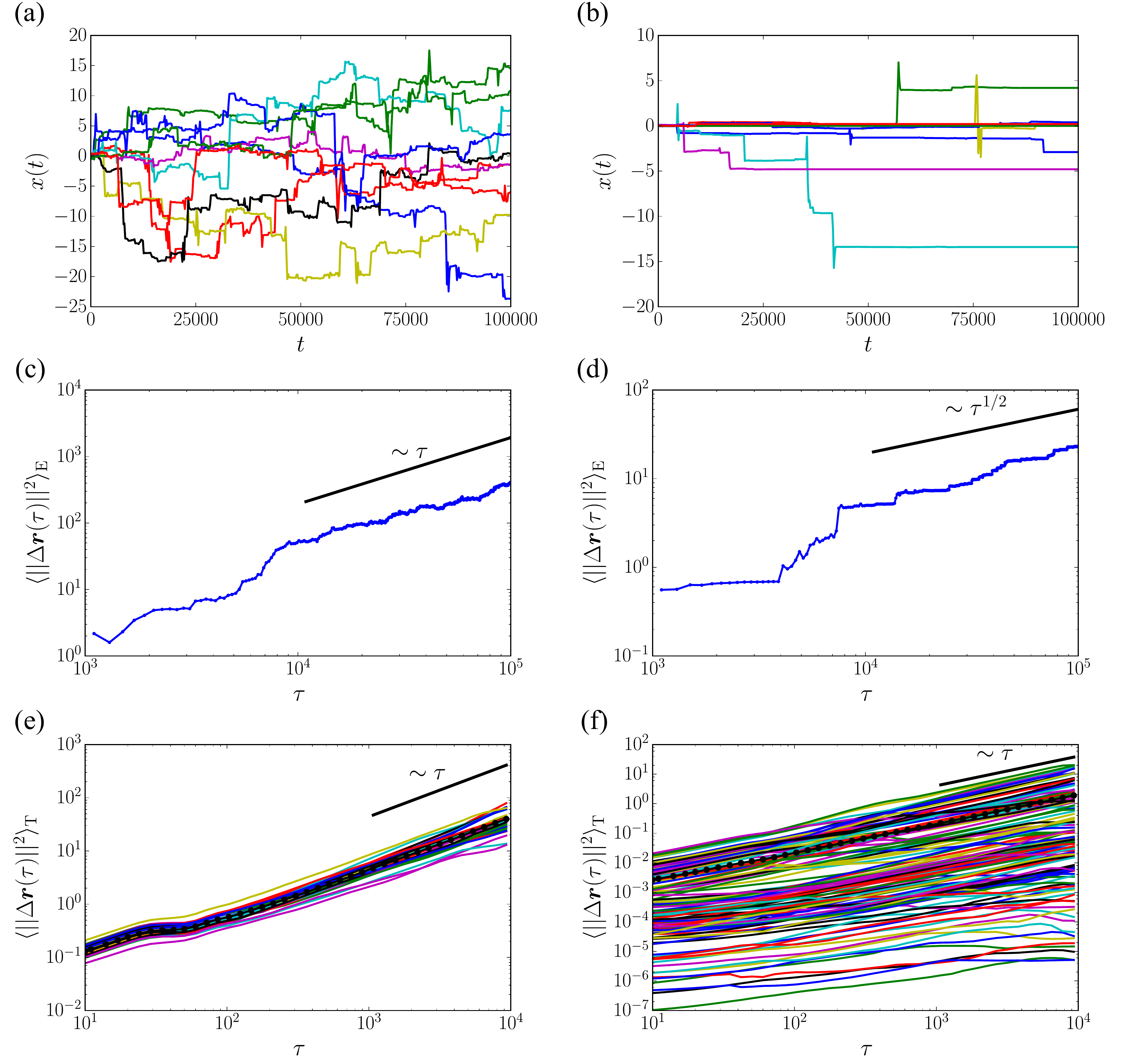}
\end{center}
\caption{Comparison of two different diffusive behaviors of the $x$-coordinate of the center of mass of solitons
that evolve in time according to Eq.~(\ref{cqgl2d}).
Left column corresponds to $\mu=-0.34$ and shows normal diffusion:
$\EASD$ grows linearly with time and $\TASD$ show up as several straight lines with similar location.
Right column corresponds to $\mu=-0.38$ and shows anomalous diffusion:
$\EASD$ grows with time following a power-law and $\TASD$ show up as several straight lines with the same slope but at very different locations.
}
\label{fig:comparison}
\end{figure*}



For $\mu=-0.340$ (left column in Fig.~\ref{fig:comparison}), all explosions are asymmetric and have independent directions.
The soliton is at rest only for short periods of time.
The resulting diffusion looks normal according to both ensemble-averaged and time-averaged squared displacements:
EASD grows proportionally with time (exponent 1 in the log-log plot),
and for the TASD, the vertical position of the curve in log-log plot (that captures the coefficient $\langle D_1 \rangle_\mathrm{T}$)
indicates that the spread between realizations is small (approaches zero for longer $T$).
As ensemble and time averages coincide, the process can be considered ergodic.

For $\mu=-0.380$ (right column in Fig.~\ref{fig:comparison}), explosions can be asymmetric or symmetric.
Long time intervals with only symmetric explosions show up as horizontal segments, 
and for large $t$, the probability of being at one of these waiting times grows.
The exponent $\alpha<1$ in the EASD captures the presence of very long waiting times. 
The greater variability of the TASD among individual trajectories also indicates anomalous diffusion.
Although the individual TASDs show linear growth with time-lag, their slopes are quite different, i.e.
the diffusion coefficient $\langle D_1 \rangle_\mathrm{T}$ is a random variable.


In Ref.~\onlinecite{CDAR16}, a more systematic way to analyze the scatter of the individual realizations
was used based on the distribution of the normalized time-averaged squared-displacement\cite{he2008}:
\begin{equation*}
\xi \mydef \frac{\TASD}{\EATASD}.
\end{equation*}
The variance of $\xi$ quantifies ergodicity breaking in the limit of $T \rightarrow \infty$.
In the case of normal diffusion, the distribution of $\xi$ is monomodal with a width that can be predicted analytically and vanishes
for long $T$.
In the case of a subdiffusive continuous time random walk, $\xi$ follows the Mittag-Leffler distribution
with a width that does not vanish for long $T$.


\section{Distribution of generalized diffusivities}

Another mathematical tool, which can be used to analyze diffusion processes, is the distribution of generalized diffusivities (DOGD)\cite{AR13,AR14,Albers}, which
captures the fluctuations of generalized diffusivities defined as scaled squared displacements,
$D_{\beta}(\tau) = ||\Delta \vect{r}(\tau)||^2/\tau^\beta$.
The advantages of the DOGD are that it obviously contains more information than the MSD,
but contrary to the propagator, it is always a one-dimensional distribution and, therefore, simple to visualize.
Furthermore, in contrast to the propagator, by choosing an appropriate value of the scaling exponent $\beta$, the DOGD may become stationary, i.e. independent of $\tau$.
Especially, if $\beta$ is equal to the exponent of the asymptotic increase of the MSD, the mean value of the DOGD is equal to the prefactor of the asymptotic increase of the MSD, i.e. the generalized diffusion coefficient.
The freedom to choose an exponent $\beta$ and a time frame $\tau$ can be used to
reveal non-trivial pieces of the process dynamics.


The distribution of generalized diffusivities obtained as ensemble average is defined by
\begin{equation}
p_\beta^\mathrm{E}(D,\tau) \mydef \left\langle \delta \left(D- \frac{||\vect{r}(\tau)-\vect{r}(0)||^2}{\tau^\beta} \right) \right\rangle_\mathrm{E},
\label{dogd_easd}
\end{equation}
where, as already mentioned, $\beta$ can be selected arbitrarily.

The distribution of generalized diffusivities obtained as time average from an individual realization of the process is defined by
\begin{equation}
p_\beta^\mathrm{T}(D,\tau) \mydef \left\langle \delta \left(D-\frac{||\vect{r}(t+\tau)-\vect{r}(t)||^2}{\tau^\beta} \right) \right\rangle_\mathrm{T}.
\label{dogd_tasd}
\end{equation}
If the generalized diffusion coefficient obtained from the TASD, which is just the first moment of the distribution $p_{\beta}^{\mathrm{T}}(D,\tau)$,
is a random variable, the distribution $p_{\beta}^{\mathrm{T}}(D,\tau)$ must be random itself.
Of course, this random distribution can be averaged over an ensemble of such distributions.

Both distributions can be computed analytically (or at least approximated by analytic expressions) in the situations that are relevant in the present context
(full details of these derivations are presented in Ref.~\onlinecite{Albers}).

\begin{itemize}
\item For a two-dimensional normal diffusion process, which is ergodic with respect to the squared displacements, we get for both definitions of the DOGD:
\begin{align}
p_\beta^\mathrm{E}(D,\tau) & = \langle p_\beta^\mathrm{T}(D,\tau) \rangle_\mathrm{E} = p_\beta(D,\tau) \nonumber \\
&= \int \delta\left( D - \frac{||\vect{r}||^2}{\tau^{\beta}}\right)\, p(\vect{r},\tau) d^2 \vect{r},
\end{align}
where
\begin{equation}
p(\vect{r},\tau) = \frac{1}{2\pi \langle D_1 \rangle \tau} \exp\left(-\frac{||\vect{r}||^2}{2 \langle D_1 \rangle\, \tau}\right)
\end{equation}
is the result for the propagator $\langle \delta (\vect{r}-\vect{r}(\tau)) \rangle_\mathrm{E}$ of the process and $\langle D_1 \rangle$ denotes the diffusion coefficient.
We obtain
\begin{equation}
p_\beta (D,\tau) = \frac{1}{2 \langle D_1 \rangle \tau^{1-\beta}} \exp\left( - \frac{D}{2\langle D_1 \rangle \tau^{1-\beta}} \right).
\label{dogd_normal}
\end{equation}
The first moment of $p_\beta(D,\tau)$ satisfies $\langle D_\beta(\tau) \rangle = \langle D_1 \rangle\, \tau^{1-\beta}$.
For the selection $\beta=1$, $p_1(D)$ does not depend on $\tau$, and
one can verify that $\langle D_1 \rangle$ is the first moment of $p_1(D)$.


\item For a subdiffusive continuous random walk characterized by exponent $\alpha<1$ and in the limit of large $\tau$, the DOGD can be computed from the propagator, which can be represented in terms of the Fox H-function\cite{SW89}:
\begin{equation}
p (\vect{r},\tau) \simeq \frac{C}{\pi} \tau^{-\alpha} H^{2,0}_{1,2} \biggl[ C \tau^{-\alpha} ||\vect{r}||^2 \biggl\vert {\scriptstyle \begin{array}{l} (1-\alpha,\alpha) \\ (0,1), (0,1) \end{array} } \biggr],
\end{equation}
where $C=[\langle D_\alpha \rangle_\mathrm{E} \Gamma(1+\alpha)]^{-1}$.
Then the DOGD obtained from an ensemble of trajectories for the selection $\beta=\alpha$ is given by
\begin{equation}
p_\alpha^\mathrm{E} (D,\tau) \simeq C H^{2,0}_{1,2} \biggl[ C D \biggl\vert {\scriptstyle \begin{array}{l} (1-\alpha,\alpha) \\ (0,1), (0,1) \end{array} } \biggr].
\label{dogd_ctrw}
\end{equation}

One can show that the distribution $p_1^\mathrm{T} (D,\tau)$ is a linear combination
of a singular part $\delta (D)$ caused by the long waiting times and a continuous part described by a Fox H-function.
The weights of the two parts are random and depend on the random generalized diffusion coefficient $\langle D_1 \rangle_\mathrm{T}$.
The ensemble average of the distribution $p_1^\mathrm{T} (D,\tau)$ then takes the form:
\begin{multline}
\langle p_1^\mathrm{T} (D,\tau)\rangle_\mathrm{E} \simeq \left( 1-\frac{1}{\Gamma(2-\alpha)\Gamma(\alpha+1)} \left(\frac{\tau}{T} \right)^{1-\alpha} \right) \delta (D) \\
+\frac{C}{\Gamma(\alpha+1)} \left(\frac{\tau}{T} \right)^{1-\alpha}
\tau^{1-\alpha} H^{2,0}_{1,2} \Biggl[ C D \tau^{1-\alpha} \Biggl\vert \begin{array}{l} (2-2\alpha,\alpha) \\ (0,1), (0,1) \end{array} \Biggr].
\label{dogd_ctrw_2}
\end{multline}

\end{itemize}

In Appendix \ref{sec:transformation}, it is shown how the Fox H-function can be transformed into a Meijer G-function. Furthermore, Appendix \ref{sec:asymptotic} provides asymptotic expansions of the Fox H-function.

\begin{figure*}
\begin{center}
\includegraphics[width=\textwidth]{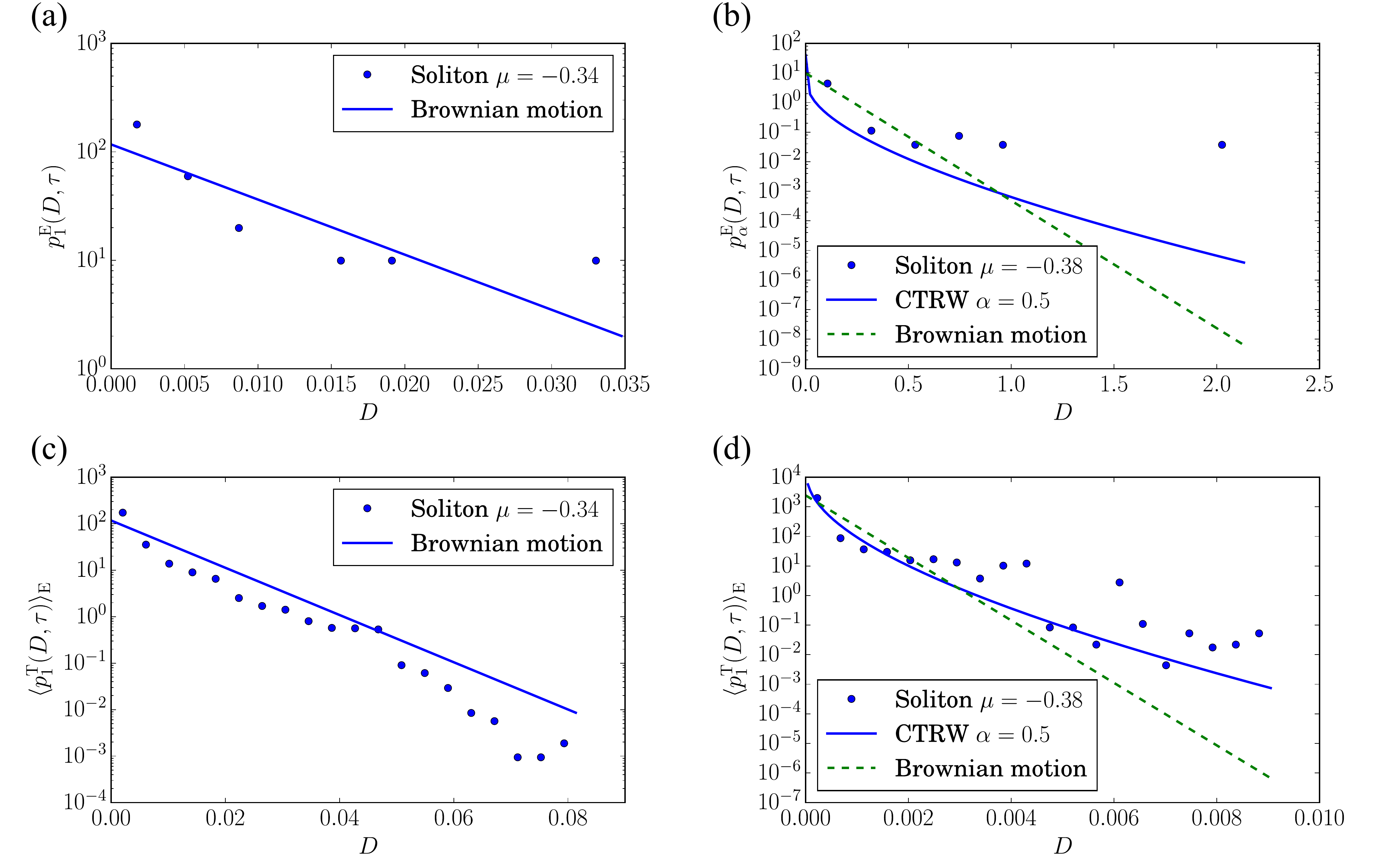}
\end{center}
\caption{Comparison of two different diffusive behaviors using the distributions of generalized diffusivities defined by Eqs.~(\ref{dogd_easd}) and (\ref{dogd_tasd})
and evaluated with soliton's data (blue points). We used $\tau=10^4, T=10^5$.
For each value of $\mu$, 64 slightly different initial conditions were used
for the integration of the soliton's trajectories.
(a,c) DOGDs for $\mu=-0.340$ (normal diffusion). Despite that the small number of realizations
prevented ensemble averages to converge, we obtained reasonable agreement with Eq.~(\ref{dogd_normal}) with $\beta=1$ deduced from Brownian motion (solid blue line).
(b,d) DOGDs for $\mu=-0.380$ (anomalous diffusion).
In (b) we obtained good agreement with Eq.~(\ref{dogd_ctrw}) deduced from subdiffusive
CTRW with exponent $\alpha=1/2$ (blue solid curve) that captures the curvature induced by anomalous diffusion.
We also show for comparison the distribution Eq.~(\ref{dogd_normal}) of Brownian motion
evaluated with $\beta=1/2$ (dashed green curve).
In (d) there is good agreement with Eq.~(\ref{dogd_ctrw_2}).
We also show for comparison the distribution Eq.~(\ref{dogd_normal}) evaluated with $\beta=1$ (dashed green curve).
}
\label{fig:dogd}
\end{figure*}

Comparisons between the DOGD from the soliton's trajectories ($\mu=0.34$ and $\mu=0.38$) and the analytical predictions
for each regime are presented in Fig.~\ref{fig:dogd}.
For $p_\beta^\mathrm{E}(D,\tau)$, we use the characteristic exponent $\beta=1$ for the normal random walk,
and $\beta=\alpha$ for the anomalous (subdiffusive) random walk.
For $\langle p_\beta^\mathrm{T}(D,\tau) \rangle_{\mathrm{E}}$, we use $\beta=1$ for both regimes.

Despite the small number of trajectories, we can appreciate how the
analytical expressions of the DOGD in both forms (and without any further adjustment of parameters)
capture two distinct features of anomalous diffusion,
a considerable probability of jumps much larger than the average and long periods of time with very little displacement.
These features are consistent with the soliton's data and provide another way of identifying
the regime.

\section{Conclusions}

In this work, we explored the erratic trajectories of two-dimensional dissipative solitons
modeled by the complex Ginzburg-Landau equation, relevant in nonlinear optics and other fields of physics.

In particular, we focused into the transition between the static soliton and the normally diffusing soliton
as one parameter of the model was varied.
The trajectories of the solitons can be described by a random walk,
but in some cases they may show features that are quite different from what one could expect
from a normal random walk or the Brownian process.
The transition is not simple as indicated by nontrivial phenomena in an intermediate range
of parameters.

In this work, we have used the ensemble-averaged squared displacement and the
time-averaged square displacement as basic mathematic tools.
The concept of the distribution of generalized diffusivities, recently introduced by some of the authors,
extended the analysis possibilities of the squared displacements
and allowed us to make comparisons between soliton's data and analytic predictions.

For the anomalous regime, the distributions of generalized diffusivities showed a direct
effect of long periods of time when the soliton exploded symmetrically and thus
remained basically static.


\begin{acknowledgments}
This work was funded in part by the Chilean Science and Technology Commission (CONICYT), grant FR-1170460.
\end{acknowledgments}

\bibliographystyle{plain}
\bibliography{buc}

\appendix 

\begin{widetext}

\section{Transformation of the Fox H-function into a Meijer G-function}
\label{sec:transformation}

In this appendix, we demonstrate with four examples how the Fox H-function for rational values
of its parameters can be transformed into the Meijer G-function (tabulated in Ref.~\onlinecite{Hfunction}
and implemented in the software \emph{Mathematica} and the Python library \emph{mpmath}),
allowing plots of our analytical results.

Our first example occurs in the context of the distribution of generalized diffusivities
$p^E_\alpha(D,\tau)$ obtained from an ensemble of subdiffusive CTRW trajectories in a two-dimensional
space for $\alpha=1/2$:
\begin{align} \label{h1}
H^{2,0}_{1,2} \Biggl[ x \Biggl\vert \begin{matrix} (\oh,\oh) \\ (0,1), (0,1) \end{matrix} \Biggr] &=
\frac{1}{2\pi\mi} \int_\mathcal{C} \frac{\Gamma(u)\Gamma(u)}{\Gamma(\oh+\frac{u}{2})}\: x^{-u}\: \rd u \nonumber \\
&= \frac{1}{4\pi} H^{3,0}_{0,3} \Biggl[ \frac{x}{4} \Biggl\vert \begin{array}{l} \text{---} \\ (0,\oh), (0,\oh), (\oh, \oh) \end{array} \Biggr] \nonumber \\
&= \frac{1}{2\pi} H^{3,0}_{0,3} \Biggl[ \left(\frac{x}{4}\right)^2 \Biggl\vert \begin{array}{l} \text{---} \\ (0,1), (0,1), (\oh, 1) \end{array} \Biggr] \nonumber \\
&= \frac{1}{2\pi} G^{3,0}_{0,3} \Biggl[ \left(\frac{x}{4}\right)^2 \Biggl\vert \begin{array}{l} \text{---} \\ 0,0, \oh \end{array} \Biggr]
\end{align}
Here, we used the definition of the Fox H-function (Eq.~(1.2) and Eq.~(1.3) in Ref.~\onlinecite{Hfunction}),
the duplication formula of the Gamma function (Eq.~(1.59) in Ref.~\onlinecite{Hfunction}),
\begin{equation*}
\Gamma(u)=\Gamma\left(\frac{u}{2}\right)\, \Gamma\left(\oh+\frac{u}{2}\right)\, \frac{2^{u-1}}{\sqrt{\pi}}
\end{equation*}
so the integrand
\begin{equation*}
\frac{\Gamma(u)\Gamma(u)}{\Gamma(\oh+\frac{u}{2})} = \Gamma\left(\frac{u}{2}\right)\, \Gamma\left(\frac{u}{2}\right)\, \Gamma\left(\oh+\frac{u}{2}\right)\, \frac{2^{2u}}{4\pi}
\end{equation*}
and the definition of the Meijer G-function (Eq.~(1.112) in Ref.~\onlinecite{Hfunction}).

The second example appears in the context of the distribution of generalized diffusivities
$p^T_1 (D,\tau)$ obtained as time average from one single-particle trajectory of a two-dimensional,
subdiffusive CTRW for $\alpha=1/2$:
\begin{align} \label{h2}
H^{2,0}_{1,2} \Biggl[ x \Biggl\vert \begin{array}{l} (1,\oh) \\ (0,1), (0,1) \end{array} \Biggr] &=
\frac{1}{2\pi\mi} \int_\mathcal{C} \frac{\Gamma(u)\Gamma(u)}{\Gamma(1+\frac{u}{2})}\: x^{-u}\: \rd u \nonumber \\
&= \frac{1}{4\pi} H^{4,0}_{1,4} \Biggl[ \frac{x}{4} \Biggl\vert \begin{array}{l} (1,\oh) \\ (0,\oh), (0,\oh), (\oh, \oh), (\oh, \oh) \end{array} \Biggr] \nonumber \\
&= \frac{1}{2\pi} H^{4,0}_{1,4} \Biggl[ \left(\frac{x}{4}\right)^2 \Biggl\vert \begin{array}{l} (1,1) \\ (0,1), (0,1), (\oh, 1), (\oh, 1) \end{array} \Biggr] \nonumber \\
&= \frac{1}{2\pi} G^{4,0}_{1,4} \Biggl[ \left(\frac{x}{4}\right)^2 \Biggl\vert \begin{array}{l} 1 \\ 0,0, \oh, \oh \end{array} \Biggr]
\end{align}
For this calculation, we used the same methods as in the first example.

The third and the fourth examples correspond to the first two examples, but now for $\alpha=3/4$:
\begin{align} \label{h3}
H^{2,0}_{1,2} \Biggl[ x \Biggl\vert \begin{array}{l} (\oq,\frac{3}{4}) \\ (0,1), (0,1) \end{array} \Biggr] &=
\frac{1}{2\pi\mi} \int_\mathcal{C} \frac{\Gamma(u)\Gamma(u)}{\Gamma(\oq+\frac{3u}{4})}\: x^{-u}\: \rd u \nonumber \\
&= \frac{\sqrt[4]{3}}{(4\pi)^2} H^{7,0}_{2,7} \Biggl[ \frac{\sqrt[4]{27}}{16}x \Biggl\vert \begin{array}{l} (\frac{1}{12}, \oq),(\frac{5}{12},\oq) \\ (0,\oq), (0,\oq), (\oq, \oq), (\oq, \oq), (\oh, \oq), (\oh, \oq), (\frac{3}{4},\oq) \end{array} \Biggr] \nonumber \\
&= \frac{\sqrt[4]{3}}{(2\pi)^2} H^{7,0}_{2,7} \Biggl[ \left(\frac{\sqrt[4]{27}}{16}x\right)^4 \Biggl\vert \begin{array}{l} (\frac{1}{12},1),(\frac{5}{12},1) \\ (0,1), (0,1), (\oq, 1), (\oq, 1), (\oh, 1), (\oh, 1), (\frac{3}{4},1) \end{array} \Biggr] \nonumber \\
&= \frac{\sqrt[4]{3}}{(2\pi)^2} G^{7,0}_{2,7} \Biggl[ \left(\frac{\sqrt[4]{27}}{16}x\right)^4 \Biggl\vert \begin{array}{l} \frac{1}{12},\frac{5}{12} \\ 0,0, \oq,\oq,\oh,\oh,\frac{3}{4} \end{array} \Biggr]
\end{align}

\begin{align} \label{h4}
H^{2,0}_{1,2} \Biggl[ x \Biggl\vert \begin{array}{l} (\oh,\frac{3}{4}) \\ (0,1), (0,1) \end{array} \Biggr] &=
\frac{1}{2\pi\mi} \int_\mathcal{C} \frac{\Gamma(u)\Gamma(u)}{\Gamma(\oh+\frac{3u}{4})}\: x^{-u}\: \rd u \nonumber \\
&= \frac{1}{(4\pi)^2} H^{7,0}_{2,7} \Biggl[ \frac{\sqrt[4]{27}}{16}x \Biggl\vert \begin{array}{l} (\frac{1}{6}, \oq),(\frac{5}{6},\oq) \\ (0,\oq), (0,\oq), (\oq, \oq), (\oq, \oq), (\oh, \oq), (\frac{3}{4}, \oq), (\frac{3}{4},\oq) \end{array} \Biggr] \nonumber \\
&= \frac{1}{(2\pi)^2} H^{7,0}_{2,7} \Biggl[ \left(\frac{\sqrt[4]{27}}{16}x\right)^4 \Biggl\vert \begin{array}{l} (\frac{1}{6},1),(\frac{5}{6},1) \\ (0,1), (0,1), (\oq, 1), (\oq, 1), (\oh, 1), (\frac{3}{4}, 1), (\frac{3}{4},1) \end{array} \Biggr] \nonumber \\
&= \frac{1}{(2\pi)^2} G^{7,0}_{2,7} \Biggl[ \left(\frac{\sqrt[4]{27}}{16}x\right)^4 \Biggl\vert \begin{array}{l} \frac{1}{6},\frac{5}{6} \\ 0,0, \oq,\oq,\oh,\frac{3}{4},\frac{3}{4} \end{array} \Biggr]
\end{align}

In this work, we fixed the parameters of Eq.~(\ref{cqgl2d})
with the exception of $\mu$, that was used to explore different scenarios.
Subdiffusive motion was found for $\mu=-0.35$ and $\mu=-0.38$ with exponents $\alpha=3/4$ and $\alpha=1/2$, respectively.
Eqs.~(\ref{h1}--\ref{h4}) can be used to compute $p^E_\alpha(D,\tau)$ and $p^T_1 (D,\tau)$ and compare with estimations based on the soliton's trajectories.

\section{Asymptotic expansions of the Fox H-function}
\label{sec:asymptotic}

In this appendix, we provide asymptotic expansions for the Fox H-functions occurring in the main text.
First, we consider the Fox H-function appearing in the formula for the distribution of generalized diffusivities $p_{\alpha}^{\text{E}}(D,\tau)$ obtained from an ensemble of two-dimensional subdiffusive CTRW trajectories.
By applying the residue theorem\cite{Hfunction}, we obtain
\begin{equation}
\begin{split}
\label{h5}
H^{2,0}_{1,2}\left[x\left|\begin{array}{l}(1-\alpha,\alpha)\\(0,1),(0,1)\end{array}\right.\right]&=\frac{1}{2\pi\text{i}}\int_{\mathcal{C}}\frac{\Gamma(u)\Gamma(u)}{\Gamma(1-\alpha+\alpha u)}x^{-u}\,\text{d}u\\
&=\sum\limits_{n=0}^{\infty}\text{Res}_{-n}\left[\frac{\Gamma(u)\Gamma(u)}{\Gamma(1-\alpha+\alpha u)}x^{-u}\right]\\
&=\sum\limits_{n=0}^{\infty}\lim\limits_{u\rightarrow-n}\frac{\text{d}}{\text{d}u}\left[(u+n)^2\frac{\Gamma(u)\Gamma(u)}{\Gamma(1-\alpha+\alpha u)}x^{-u}\right]\\
&=\sum\limits_{n=0}^{\infty}\frac{2(H_n-\gamma)-\alpha\psi(1-\alpha-\alpha n)-\log(x)}{\Gamma(1-\alpha-\alpha n)}\frac{x^n}{(n!)^2},
\end{split}
\end{equation}
where $H_n=\sum\limits_{k=1}^n\frac{1}{k}$ is the harmonic number, $\gamma=0.5772...$ denotes Euler's constant, and $\psi(x)=\Gamma'(x)/\Gamma(x)$ is the digamma function.
By truncating the infinite series in Eq.~(\ref{h5}) after a finite number of terms, we obtain the asymptotic behavior of the Fox H-function for $x\rightarrow0$.
Correspondingly, for the Fox H-function occurring in the context of the distribution of generalized diffusivities $p_1^{\text{T}}(D,\tau)$ obtained as time average from one single-particle trajectory, we get
\begin{equation}
\label{h6}
H^{2,0}_{1,2}\left[x\left|\begin{array}{l}(2-2\alpha,\alpha)\\(0,1),(0,1)\end{array}\right.\right]=\sum\limits_{n=0}^{\infty}\frac{2(H_n-\gamma)-\alpha\psi(2-2\alpha-\alpha n)-\log(x)}{\Gamma(2-2\alpha-\alpha n)}\frac{x^n}{(n!)^2}.
\end{equation}
The asymptotic behavior of the Fox H-functions for $x\rightarrow\infty$ can be determined by applying theorem 4 on page 289 in Ref.~\onlinecite{braaksma1963}.
We obtain
\begin{equation}
\label{h7}
H^{2,0}_{1,2}\left[x\left|\begin{array}{l}(1-\alpha,\alpha)\\(0,1),(0,1)\end{array}\right.\right]\simeq(2-\alpha)^{-\frac{1}{2}}\alpha^{\alpha-\frac{1}{2}}\left(\alpha^{\alpha}x\right)^{\frac{\alpha-1}{2-\alpha}}\exp\left[-(2-\alpha)\alpha^{\frac{\alpha}{2-\alpha}}x^{\frac{1}{2-\alpha}}\right]\quad(x\rightarrow\infty)
\end{equation}
and
\begin{equation}
\label{h8}
H^{2,0}_{1,2}\left[x\left|\begin{array}{l}(2-2\alpha,\alpha)\\(0,1),(0,1)\end{array}\right.\right]\simeq(2-\alpha)^{-\frac{1}{2}}\alpha^{2\alpha-\frac{3}{2}}\left(\alpha^{\alpha}x\right)^{\frac{2\alpha-2}{2-\alpha}}\exp\left[-(2-\alpha)\alpha^{\frac{\alpha}{2-\alpha}}x^{\frac{1}{2-\alpha}}\right]\quad(x\rightarrow\infty).
\end{equation}

\end{widetext}

\end{document}